\documentclass[a4paper,onecolumn,11pt,accepted=2024-01-16]{quantumarticle}
\pdfoutput=1
\usepackage{fullpage}
\usepackage[pagebackref=true,colorlinks]{hyperref}
\usepackage{amsmath,amssymb,amsthm,mathrsfs}
\usepackage{dsfont}
\usepackage[bottom]{footmisc}
\usepackage{graphicx}
\usepackage{caption}
\usepackage{enumerate}
\usepackage{framed}
\usepackage{authblk}
\hypersetup{
	colorlinks=true,
    linktoc=all,
    linkcolor=blue,  
    citecolor=blue,
    }

\newtheorem{theorem}{Theorem}

\newtheorem{corollary}{Corollary}[theorem]
\newtheorem{definition}{Definition}

\theoremstyle{definition}

\newtheorem{remark}{Remark}

\newcommand{\<}{\langle}
\renewcommand{\>}{\rangle}
\newcommand{\id}{\operatorname{id}}

\newcommand{\Tr}[1]{\operatorname{Tr}\!\left[#1\right]}

\def\id{\mathsf{id}}
\def\mE{\mathcal{E}}

\def\mL{\mathcal{L}}

\def\mR{\mathscr{R}}
\def\sH{\mathscr{H}}

\def\openone{\mathds{1}}
\newcommand{\set}[1]{\mathcal{#1}}
\newcommand{\op}[1]{\mathbb{#1}}

\renewcommand{\ge}{\geqslant}
\renewcommand{\le}{\leqslant}

\newcommand{\ket}[1]{|#1\rangle }
\newcommand{\ketbra}[1]{|#1\rangle \langle #1 |}
\newcommand{\mRtun}{\mathscr{R}_{\mathcal{X}}^{\;\operatorname{t}}}
\newcommand{\povm}[1]{\boldsymbol{\mathsf{#1}}}
\newcommand{\vecp}{\boldsymbol{p}}
\newcommand{\vecq}{\boldsymbol{q}}

\newcommand{\succsharp}{\succeq_{\set{X}}^{\operatorname{sharp}}}
\newcommand{\succcorr}{\succeq_{\set{X}}^{\operatorname{t}}}
\newcommand{\succlpsr}{\succeq_{\set{X}}^{\operatorname{LPSR}}}

\begin{document}

\title{A complete and operational resource theory of measurement sharpness}

\author{Francesco Buscemi}
\email{buscemi@i.nagoya-u.ac.jp}
\affiliation{Department of Mathematical Informatics, Nagoya University, Furo-cho, Chikusa-ku, 464-8601 Nagoya, Japan}
\orcid{0000-0001-9741-0628}
\author{Kodai Kobayashi}
\affiliation{Department of Mathematical Informatics, Nagoya University, Furo-cho, Chikusa-ku, 464-8601 Nagoya, Japan}
\author{Shintaro Minagawa}
\affiliation{Department of Mathematical Informatics, Nagoya University, Furo-cho, Chikusa-ku, 464-8601 Nagoya, Japan}
\orcid{0000-0002-8637-629X}

\begin{abstract}
    We construct a resource theory of \emph{sharpness} for finite-dimensional positive operator-valued measures (POVMs), where the \emph{sharpness-non-increasing} operations are given by quantum preprocessing channels and convex mixtures with POVMs whose elements are all proportional to the identity operator. As required for a sound resource theory of sharpness, we show that our theory has maximal (i.e., sharp) elements, which are all equivalent, and coincide with the set of POVMs that admit a repeatable measurement. Among the maximal elements, conventional non-degenerate observables are characterized as the canonical ones. More generally, we quantify sharpness in terms of a class of monotones, expressed as the EPR--Ozawa correlations between the given POVM and an arbitrary reference POVM. We show that one POVM can be transformed into another by means of a sharpness-non-increasing operation if and only if the former is sharper than the latter with respect to all monotones. Thus, our resource theory of sharpness is \emph{complete}, in the sense that the comparison of all monotones provides a necessary and sufficient condition for the existence of a sharpness-non-increasing operation between two POVMs, and \emph{operational}, in the sense that all monotones are in principle experimentally accessible.
\end{abstract}

\section{Introduction}

While quantum theory has been traditionally developed around the concept of observables as self-adjoint operators and their spectral decomposition~\cite{von1955mathematical}, it is a well-known fact that many fundamental problems, such as optimal approximate joint measurements of noncommuting observables, and applications, such as optimal parameter estimation (including quantum state tomography~\cite{paris-rehacek-quantum-state-estimation} and several instances of the state discrimination problem~\cite{bergou-2010-discrimination-states-review,dallarno-guesswork-2022}), require a more general formalism, where orthogonal projective decompositions of the identity are replaced by positive operator-valued measures, i.e., POVMs~\cite{Davies-Lewis-1970aa-CMP-operational,ozawa1980optimal,busch-lahti-mittelstaedt}.

If POVMs constitute a notion of ``approximate'' observables, it is a natural question to ask, given a POVM, how close that is to an observable. This question has led several researchers to consider the problem of formalizing a concept of ``sharpness'' as a way to provide a quantitative measure of how close a given POVM is to a proper observable~\cite{Carmeli-heinonen-toigo_2007-intrinsic-unsharpness,massar-2007-uncertainty-relations-for-POVMs,Busch-2009a-on-the-sharpness-and-bias,Baek-Son-2016aa-unsharpness,Liu-Luo-2021-PRA-quantifying-unsharpness}, where the latter is of course taken as the prototype of a perfectly sharp measurement. In a similar vein, Refs.~\cite{oszmaniec-2017-simulating-povms-with-projective,oszmaniec-2019-simulating-all-POVM-with-PVM-postselection} consider the closely related problem of how effectively general POVMs can be simulated using only perfectly sharp measurements.

Notwithstanding the fundamental and practical relevance of POVMs, conventional observables and their orthogonal projective decompositions still constitute the \textit{de facto} standard model for quantum measurements in various areas of physics, especially in quantum statistical mechanics, where ``quantum measurement'' is often used as a synonym for ``projective measurement of an observable''. The reasons for this are arguably two-fold. On the one hand, projective measurements are easier to treat mathematically than general POVMs, as the former can be conveniently represented using a single self-adjoint operator, from which many important properties can be readily determined. On the other hand, projective measurements inherently possess the operational property of \textit{repeatability}~\cite{von1955mathematical}: since projections are idempotent operators, as the name itself suggests, a projective measurement repeated twice in succession always produces the same result, which thus acquires a character of ``objectivity''\footnote{One can argue~\cite{Ozawa-2015-Heisenberg-repeat} that in the early days of quantum theory, repeatability was considered such an obvious and natural requirement that it was often implicitly assumed. The situation changed only in the 1970s and 1980s, with the establishment of the general theory of quantum measurements~\cite{Davies-Lewis-1970aa-CMP-operational,ozawa1984quantum}.}.

The reasons given above, while compelling, provide only an intuitive understanding of what makes conventional observables ``special'' among general POVMs. A promising way to make the discussion more rigorous is to try to characterize measurement sharpness as a \textit{resource}, following recent developments in quantum information theory~\cite{Chitambar-Gour2019resource-theories}. However, attempts to construct a comprehensive resource theory of sharpness have until now been only partially successful~\cite{Busch-2009a-on-the-sharpness-and-bias,Mitra-2022aa-IJTP-quantifying-unsharpness}.

In this paper, we fill this gap by proposing a complete and operational resource theory of sharpness. The picture that we obtain is that conventional non-degenerate observables can indeed be singled out as the ``canonical'' elements in the class of maximally sharp POVMs, which are identified with POVMs that admit a repeatable measurement. Below these, we find many degrees of sharpness, quantified by a \textit{robustness-like function}, so that POVMs can be ordered according to their degree of sharpness. We also introduce a class of \textit{sharpness-non-increasing operations} that can be used to transform a sharper measurement into a less sharp one. This is exactly what one would expect from a resource theory of sharpness. However, our sharpness resource theory has some additional desirable features that provide connections to several areas of independent interest. First of all, the sharpness measures (in jargon, the \emph{monotones}) that we introduce are defined using Ozawa's degrees of measurements correlation~\cite{OZAWA-2005-PLA-perfect-correlations,OZAWA-2006-Annals-Quantum-perfect-correlations} and, as such, are all in principle experimentally accessible. Second, it is shown that the class of sharpness-non-increasing transformations (in jargon, the class of \emph{free operations}) corresponds to a restricted class of preprocessing channels applied not to the given POVM, but to an extended object representing a \emph{family} of POVMs, thus establishing a direct connection with the theory of programmable measurement devices~\cite{buscemi2020complete,ji2021incompatibility,buscemi-2022-unifying-instrument-incompatibility}. Third, our sharpness monotones provide a \emph{complete comparison} (in the sense of Blackwell~\cite{blackwell1953,buscemi-CMP-2012}), in that one POVM is sharper than another with respect to all such monotones if and only if the former can be transformed into the latter by means of an appropriate free operation.

The paper is organized as follows. In Section~\ref{sec:POVM-and-preorders} we introduce notations and basic definitions, and review the pre- and postprocessing preorders of POVMs~\cite{buscemi-2005-clean-POVMs}. In Section~\ref{sec:fuzzifying} we define the set of free operations and show that they can be regarded as preprocessings on objects that extend POVMs to programmable devices. In Section~\ref{sec:ozawa-correlations-and-sharpness} we review the theory of EPR--Ozawa measurement correlations~\cite{OZAWA-2005-PLA-perfect-correlations,OZAWA-2006-Annals-Quantum-perfect-correlations} and use it to define a class of sharpness measures, which are by construction non-increasing under free operations. In Section~\ref{sec:blackwell} we prove a Blackwell--like theorem for sharpness, stating the equivalence between the sharpness preorder, arising from comparing all sharpness measures for a pair of POVMs, and the existence of a sharpness-non-increasing transformation from one POVM into the other one. Finally, in Section~\ref{sec:conclusion} we conclude the paper with a summary of our resource theory of sharpness.

\section{Quantum measurements and preorders}\label{sec:POVM-and-preorders}

Let us consider a quantum system $A$ associated with a finite $d_A$-dimensional Hilbert space $\sH_A$. Here and in what follows, all sets are considered finite and all spaces finite-dimensional. States of $A$ are in one-to-one correspondence with \emph{density matrices} on $\sH_A$, i.e., matrices $\rho_A\ge 0$ such that $\Tr{\rho_A}=1$. A quantum state $\rho_A$ contains all the information needed to predict the statistics of any observation done on it, as modeled by a positive operator-valued measure (\emph{POVM}), namely, a family $\povm{P}=\{P^x_A\}_{x\in\set{X}}$ of positive semi-definite operators $P^x_A\ge 0$ labeled by the outcome set $\set{X}$ (also assumed to be finite), such that the \emph{completeness relation} $\sum_{x\in\set{X}}P^x_A=\openone_A$, where $\openone_A$ denotes the identity operator on $\sH_A$, is satisfied. For notational convenience, we will sometimes simply take the outcome set $\set{X}$ to be a subset of the natural numbers, i.e., $\set{X}=\{1,2,\dots,N\}$.

The interpretation of POVMs in terms of quantum measurements is based on the Born rule, which postulates that to any observation with outcomes in set $\set{X}$, there corresponds a POVM so that, if the state of the system is given by $\rho_A$, the expected probability of occurrence of each outcome $x\in\set{X}$ is computed as $\Tr{P^x_A\ \rho_A}$. Notice that, in general, a POVM may contain some null elements, corresponding to the situation in which some outcomes in $\set{X}$ never occur. Further, a POVM is said to be
\begin{itemize}
    \item \emph{rank-one}, whenever all its elements $P^x_A$ are proportional to rank-one projectors;
    \item \textit{projective}, whenever all its elements $P^x_A$ are orthogonal projectors, i.e., $P^x_AP^{x'}_A=\delta_{x,x'}P^x_A$;
    \item \textit{trivial}, whenever all its elements $P^x_A$ are proportional to the identity matrix.
\end{itemize}

Any POVM on $\sH_A$ with outcome set $\set{X}$ can also be understood as a linear map from the set of density matrices on $\sH_A$ to the set of normalized probability distributions on $\set{X}$. More generally, in operational quantum theory a crucial role is played by completely positive trace-preserving (CPTP) linear maps, also known as \emph{quantum channels}, that is, linear maps transforming density matrices on an input space $\sH_A$ to density matrices on an output space $\sH_B$, in such a way that parallel compositions are well-defined\footnote{In particular, the notion of \emph{complete positivity} is required so that a quantum channel $\mE$ remains a quantum channel even if it is extended with the identity map as $\mE\otimes\id$, for any ancillary system.}. We will denote any such a quantum channel as $\mE:A\to B$ for short. The Born rule naturally associates to any quantum channel $\mE:A\to B$ a \emph{trace-dual channel} $\mE^\dag:B\to A$, which maps POVMs on $\sH_B$ to POVMs on $\sH_A$ and is defined by the equality $\Tr{\mE^\dag(Y)\ X}:=\Tr{Y\ \mE(X)}$, for all linear operators $Y$ on $\sH_B$ and $X$ on $\sH_A$. It is easy to verify that a linear map $\mE:A\to B$ is completely positive and trace-preserving if and only if its trace-dual $\mE^\dag:B\to A$ is completely positive and unit-preserving, i.e., $\mE^\dag(\openone_B)=\openone_A$.

In this paper, we propose the following definition.

\begin{definition}[sharp POVMs]\label{def:sharp-POVMs}
A POVM $\povm{P}=\{P^x_A\}_{x\in\set{X}}$ is called \emph{sharp}, whenever all its elements contain the real unit among their eigenvalues, i.e., there exist normalized vectors $|\psi^x\>_A$ such that $P^x_A|\psi^x\>_A=|\psi^x\>_A$ for all $x\in\set{X}$. POVMs that are not sharp are called \emph{unsharp} or \emph{fuzzy}.
\end{definition}

Notice that our definition of sharp POVMs differs from the usual one~\cite{Busch-2009a-on-the-sharpness-and-bias,Liu-Luo-2021-PRA-quantifying-unsharpness,Mitra-2022aa-IJTP-quantifying-unsharpness}, where sharp POVMs are identified with projective POVMs. For example, while it is obvious that a sharp POVM need not be projective, also a projective POVM may be not sharp (according to our definition) if it contains zero elements. However, sharp POVMs share with projective POVMs the operational property of being measurable in a repeatable way\footnote{With the difference that the corresponding repeatable measurement may be not of the von Neumann--L\"uders (or ``square-root'') type~\cite{von1955mathematical,loeders1951uber}, but rather of the Gordon--Louisell (or ``measure-and-prepare'') type~\cite{gordon-louisell}. In order to discuss further the notion of repeatability one should employ the concept of \emph{quantum instruments}, but this point is beyond the scope of the present paper. We refer the interested reader to~\cite{ozawa1984quantum,busch-grabowski-lahti-1995-operational,buscemi2004repeatable} for a careful presentation of the problem and some fundamental results.}. In fact, except for the trivial case of outcomes that are repeatable simply because they never occur (i.e., those outcomes corresponding to null POVM elements), it is sharpness (as defined here) and not projectivity that is the main ingredient for repeatability~\cite[Section II.3.5]{busch-grabowski-lahti-1995-operational}. For example, a simple way to measure a sharp POVM in a repeatable way might be to prepare the system in the pure state $|\psi^x\>_A$ whenever outcome $x$ is observed: in this way, it is clear that a subsequent measurement of the same POVM would always yield the same result.

Notice also that the completeness relation $\sum_xP^x_A=\openone_A$ implies that the cardinality of the outcome set of a sharp POVM cannot exceed the dimension of the underlying Hilbert space, i.e., $|\set{X}|\le d_A$. Hence, among all sharp POVMs with outcome set $\set{X}$, the ``smallest'' or \textit{canonical} ones, i.e., those with $d_A=|\set{X}|$, exactly coincide with rank-one projective POVMs, i.e., conventional non-degenerate observables.

Another consequence of the completeness relation is that, for a sharp POVM, it must also be the case that $P^{x'}_A|\psi^x\>_A=0$ for all $x'\neq x$, i.e., the elements of a sharp POVM can be perfectly discriminated, thus leading to the maximum \textit{informational power} for any given outcome set $\set{X}$~\cite{dallarno-2011-info-power,dallarno-buscemi-ozawa-2014-info-power-bounds}. In other words, a POVM $\povm{P}=\{P^x_A\}_{x\in\set{X}}$ is sharp if and only if the set of all probability distributions on $\set{X}$ that can be obtained from $\povm{P}$ by varying the state $\rho_A$ of the system, i.e., its \textit{testing region}~\cite{buscemi-gour-2017-q-lorenz,dallarno-buscemi-2023-testing-regions}
\begin{align*}
    \mathcal{P}(\povm{P}):=\{\vecp\in\mathbb{R}^{|\set{X}|}:\exists\rho_A \text{ s.t. }p_x=\Tr{P^x_A\;\rho_A}\forall x\in\set{X}\}\;,
\end{align*}
coincides with the entire probability simplex on $\set{X}$.

\subsection{Preorders of POVMs}

Given a set $\Omega=\{\omega:\omega\in\Omega\}$, a preorder is a binary relation $\succeq$ between elements of $\Omega$ that is reflexive (i.e., $\omega\succeq\omega$) and transitive (i.e., if $\omega\succeq\omega'$ and $\omega'\succeq\omega''$, then $\omega\succeq\omega''$). An element $\omega$ is \textit{maximal} if there exists no other element strictly above $\omega$: that is, for any other element $\omega'$ with $\omega'\succeq\omega$, then also $\omega\succeq\omega'$ holds. Analogously, an element an element $\omega$ is \textit{minimal} if there exists no other element strictly below $\omega$: that is, for any other element $\omega'$ with $\omega\succeq\omega'$, then also $\omega'\succeq\omega$ holds.

Two preorders that are relevant for the study of the mathematical properties of POVMs, including their sharpness, are the \emph{quantum preprocessing preorder} and the \emph{classical postprocessing preorder}~\cite{buscemi-2005-clean-POVMs}, which are defined as follows.

Given two POVMs $\povm{P}=\{P^x_A\}_{x\in\set{X}}$ and $\povm{Q}=\{Q^x_B\}_{x\in\set{X}}$, possibly defined on different Hilbert spaces $\sH_A$ and $\sH_B$ but with the same outcome set $\set{X}$, we say that $\povm{P}$ is \emph{preprocessing cleaner} than $\povm{Q}$,
whenever there exists a quantum channel $\mE:B\to A$ such that $\mE^\dag(P^x_A)=Q^x_B$ for all $x\in\set{X}$.

Further, given two POVMs $\povm{P}=\{P^x_A\}_{x\in\set{X}}$ and $\povm{Q}=\{Q^y_A\}_{y\in\set{Y}}$, possibly with different outcome sets but defined on the same Hilbert space $\sH_A$, we say that $\povm{P}$ is \emph{postprocessing cleaner} than $\povm{Q}$,
whenever there exists a conditional probability distribution $\mu(y|x)$ such that $Q^y_A=\sum_x\mu(y|x)P^x_A$ for all $y\in\set{Y}$.

\begin{remark}
Notice that a necessary condition for $\povm{P}$ to be preprocessing cleaner than $\povm{Q}$ is that if, for some $x\in\set{X}$, $P^x_A=0$, then also $Q^x_B=0$. That is, outcomes that never occur for $\povm{P}$ cannot occur for $\povm{Q}$ either. This is a consequence of the fact that $\mE^\dag$ is linear element-wise, that is, on each POVM element. Instead, the classical postprocessing preorder is more flexible: for example, it may swap an outcome corresponding to a null POVM element with an outcome corresponding to a non-zero operator.
\end{remark}

The connection between sharpness and POVMs preorders arises from the following result proved in Ref.~\cite{buscemi-2005-clean-POVMs}.

\begin{theorem}\label{th:pre-clean}
A POVM $\povm{P}=\{P^x_A\}_{x\in\set{X}}$, defined on a Hilbert space $\sH_A$ and with outcome set $\set{X}$, is sharp if and only if $\povm{P}$ is preprocessing cleaner than any other POVM with the same outcome set $\set{X}$.
\end{theorem}

\begin{remark}\label{rem:maximally-fuzzy}
    Theorem~\ref{th:pre-clean} identifies sharp POVMs as the \textit{maximal} elements of the quantum preprocessing preorder. Instead, the minimal (or ``maximally fuzzy'') elements turn out to be the trivial POVMs\footnote{When the outcome set $\set{X}$ is a singleton, there exists only one POVM, which is at once maximal and minimal, sharp and trivial.}. This is due to the fact that the map $\mE^\dag$ is linear and unit-preserving, so that any trivial POVM on $A$, such as $\{p(x)\openone_A\}_{x\in\set{X}}$ for some probability distribution $p(x)$, cannot be transformed into anything that is not trivial. In fact, any trivial POVM cannot be transformed into anything but itself (apart from changing the system). This fact constitutes an important difference between the maximal and the minimal elements of the quantum preprocessing preorder. While the former are all equivalent, the latter are all \textit{inequivalent}. Since in a sound resource theory ``free objects'' are all equivalent (simply because they can be freely generated, by definition), we understand that a resource theory of sharpness should include more general operations than just quantum preprocessing channels. We will return to this point in the next section.
\end{remark}

\begin{proof}[Proof of Theorem~\ref{th:pre-clean}]
We briefly recount here the proof of the above theorem for the sake of completeness. We begin by showing that any sharp POVM is a maximal element for the quantum preprocessing preorder. If a POVM $\povm{P}=\{P^x_A\}_{x\in\set{X}}$ is sharp, then there exist normalized vectors $|\psi^x\>_A$ such that $P^x_A|\psi^{x'}\>_A=\delta_{x,x'}|\psi^{x'}\>_A$. This condition in particular implies that the normalized vectors $|\psi^x\>_A$ are also mutually orthogonal. Consider then the linear operator from $\sH_B$ to $\sH_A\otimes\sH_B$
\begin{align*}
V:=\sum_{x\in\set{X}}|\psi^x\>_A\otimes\sqrt{Q^x_B}\;,
\end{align*}
which takes a vector $|\varphi\>_B$ in $\sH_B$ to $\sum_{x\in\set{X}}|\psi^x\>_A\otimes\sqrt{Q^x_B}|\varphi\>_B$ in $\sH_A\otimes\sH_B$.
Notice that it may be that, for some $x$, $Q^x_B=0$. (Instead, the POVM $\povm{P}$ is assumed to be sharp.) It is easy to check that $V$ is an isometry, since $V^\dag V=\sum_xQ^x_B=\openone_B$. Moreover, by direct inspection,
\begin{align*}
V^\dag (P^x_A\otimes\openone_B)V=Q^x_B\;,
\end{align*}
for all $x\in\set{X}$. Since the linear map $V^\dag(\bullet_A\otimes\openone_B)V$ is by construction completely positive and identity-preserving, the above equation shows that $\povm{P}$ is preprocessing cleaner than $\povm{Q}$, for any $\povm{Q}$, as claimed.

Conversely, let us suppose that $\povm{P}$ is preprocessing cleaner than $\povm{Q}$, for any other POVM $\povm{Q}$. This is equivalent to say that, however we choose the POVM elements $Q^x_B$, there exists a completely positive unit-preserving linear map $\mE^\dag:A\to B$ such that $\mE^\dag(P^x_A)=Q^x_B$ for all $x\in\set{X}$. Let then $Q^x_B$ constitute a sharp POVM, that is, all $Q^x_B$'s have the real number one as an eigenvalue. Now we invoke the fact that a completely positive unit-preserving linear map is spectrum-width decreasing: putting $|\varphi^x\>_B$ such that $Q^x_B|\varphi^x\>_B=|\varphi^x\>_B$ for all $x$, we have
\begin{align*}
1=\Tr{\ketbra{\varphi^x}_B\;Q^x_B}=\Tr{\ketbra{\varphi^x}_B\;\mE^\dag(P^x_A)}=\Tr{\mE(\ketbra{\varphi^x}_B)\;P^x_A}\le 1\;,
\end{align*}
that is, the real unit must already be an eigenvalue also of all $P^x_A$'s, whence their sharpness.
\end{proof}

Instead, the classical postprocessing preorder is not related with the sharpness of POVMs, but rather to their being rank-one or not~\cite{martens1990nonideal,buscemi-2005-clean-POVMs}. For example, a POVM with repeated elements like the following
\begin{align*}
\left\{\frac12\ketbra{\psi^1}_A,\;\frac12\ketbra{\psi^1}_A,\;\frac12\ketbra{\psi^2}_A,\;\frac12\ketbra{\psi^2}_A,\dots \right\}\;,        
\end{align*}
is postprocessing clean (because it is rank-one), even though it is obviously unsharp. Nonetheless, assuming $\<\psi^i|\psi^j\>=\delta_{ij}$, the POVM above can be turned into a sharp POVM by classically postprocessing its element: merging the outcomes two by two, we obtain the projective (and thus, sharp) POVM $\left\{\ketbra{\psi^1}_A,\;\ketbra{\psi^2}_A,\dots \right\}$. This arguably is the reason why attempts to characterize POVMs sharpness using classical postprocessings can only be partially successful, as noticed in~\cite{Mitra-2022aa-IJTP-quantifying-unsharpness}.

\section{Fuzzifying operations}\label{sec:fuzzifying}

In the light of Theorem~\ref{th:pre-clean} and Remark~\ref{rem:maximally-fuzzy}, it is tempting to conclude that \textit{sharpness-non-increasing} or \textit{fuzzifying} operations exactly coincide with quantum preprocessing channels. This however cannot be the case for a resource theory of sharpness, as the following example shows. Let us consider two trivial POVMs, such as $\{0,\openone\}$ and $\{\openone,0\}$. Since any quantum preprocessing channel is linear and unit-preserving, as noticed in Remark~\ref{rem:maximally-fuzzy}, it is impossible to transform $\{0,\openone\}$ into $\{\openone,0\}$ or vice versa, since both $0$ and $\openone$ are fixed points for any linear unit-preserving map. In fact, any trivial POVM on system $A$ can only be transformed into the corresponding trivial POVM on $B$. This simple observation leads us to conclude that, if free operations were given only by quantum preprocessing channels, the resulting resource theory would have many inequivalent minimal, i.e., resource-free, elements. Instead, one would like a resource theory of sharpness to have \textit{all} trivial POVMs \textit{equivalent} to each other, following the prescription that resource-free objects in a resource theory should all be freely available under free operations~\cite{Chitambar-Gour2019resource-theories}.

We thus introduce the following definition:

 \begin{definition}[sharpness preorder]\label{def:sharpness-preorder}
Given two POVMs $\povm{P}=\{P^x_A\}_{x\in\set{X}}$ and $\povm{Q}=\{Q^x_B\}_{x\in\set{X}}$, possibly defined on different Hilbert spaces $\sH_A$ and $\sH_B$ but with the same outcome set $\set{X}$, we say that $\povm{P}$ is \emph{sharper} than $\povm{Q}$, and write
\begin{align}
\povm{P}\succsharp\povm{Q}\;,
\end{align}
whenever there exists a quantum channel $\mE:B\to A$, a trivial POVM $\{p(x)\openone_B\}_{x\in\set{X}}$ on $B$, and a number $\mu\in[0,1]$, such that
\begin{align*}
    Q^x_B=\mu\mE^\dag(P^x_A)+(1-\mu)p(x)\openone_B\;,
\end{align*}
for all $x\in\set{X}$.    
\end{definition}
The above definition immediately suggests also the following:
\begin{definition}[fuzzifying operations]\label{def:fuzzifying-op}
    Given a POVM $\povm{P}=\{P^x_A\}_{x\in\set{X}}$, a fuzzifying operation on $\povm{P}$ is any transformation of the form
\begin{align}\label{eq:def-fuzzifying}
\forall x\in\set{X}\;,\qquad    P^x_A\mapsto\mu\mE^\dag(P^x_A)+(1-\mu)p(x)\openone_B\;,
\end{align}
for some arbitrary but fixed probability $\mu\in[0,1]$, probability distribution $p(x)$, and quantum preprocessing channel $\mE^\dag:A\to B$.
\end{definition}

Then, Definition~\ref{def:sharpness-preorder} can be reformulated as follows: $\povm{P}\succsharp\povm{Q}$ if and only if there exists a fuzzifying operation transforming $\povm{P}$ into $\povm{Q}$. It is easy to see that the maximal elements of $\succsharp$ are all equivalent and coincide with sharp POVMs, as it was the case for the quantum preprocessing preorder. Now, however, also all trivial POVMs turn out to be equivalent to each other, thus solving the problem that we raised in Remark~\ref{rem:maximally-fuzzy}.

But at this point another problem arises: fuzzifying operations, seen as maps acting on the POVM elements $P^x_A$, are in general \emph{not} linear, since they could transform zero operators into non-zero operators. But neither they are combinations of quantum preprocessing and classical postprocessing\footnote{In fact, as noticed in Ref.~\cite{Mitra-2022aa-IJTP-quantifying-unsharpness}, classical postprocessing can increase sharpness, and thus cannot be part of sharpness-non-increasing operations.} of the POVM $\povm{P}$. Thus, the question is how fuzzifying operations can be understood \emph{operationally}.

\subsection{Local preprocessing with shared randomness (LPSR)}

The starting point is to reformulate Definition~\ref{def:sharpness-preorder} as follows: denoting by $\povm{T}^{(i)}=\{T^{x|i}_B\}_{x\in\set{X}}$ the extremal trivial POVM on $B$ such that $T^{x|i}_B=\delta_{x,i}\openone_B$ for all $x,i\in\set{X}$, then $\povm{P}$ is sharper than $\povm{Q}$ if and only if $\povm{Q}$ belongs to the convex hull of $\{\mE^\dag(\povm{P})\}\cup\{\povm{T}^{(i)}\}_{i\in\set{X}}$. This suggests the following construction.

Given a finite outcome set $\set{X}=\{1,2,\dots,N\}$, we take as the \textit{objects} of the theory not just POVMs with outcome set $\set{X}$, but rather \textit{families} comprising $N+1$ POVMs: the first POVM, which is the given POVM $\povm{P}$ whose sharpness is being evaluated, together with the $N$ extremal trivial POVMs $\povm{T}^{(i)}$, with $i=1,\dots,N$, introduced above. More explicitly, given a POVM $\povm{P}=\{P^x_A\}_{x\in\set{X}}$, the corresponding object in the resource theory is given by the family
\begin{align}
    \overline{\povm{P}}&\equiv\{\povm{P}_0,\povm{P}_1,\povm{P}_2,\dots,\povm{P}_N\}\nonumber\\
    &:=\{\povm{P},\povm{T}^{(1)},\povm{T}^{(2)},\dots\povm{T}^{(N)}\}\label{eq:extended-POVM}\\
    &=\left\{\begin{pmatrix}P^1_A\\P^2_A\\\vdots\\ P^N_A\end{pmatrix},\begin{pmatrix}\openone_A\\0\\\vdots\\ 0\end{pmatrix},\begin{pmatrix}0\\\openone_A\\\vdots\\ 0\end{pmatrix},\dots,\begin{pmatrix}0\\0\\\vdots\\ \openone_A\end{pmatrix}\right\}\;.\nonumber
\end{align}
Notice that there is a one-to-one correspondence between POVMs $\povm{P}$ and extended families $\overline{\povm{P}}$. Hence, in what follows, when writing $\overline{\povm{P}}$ we will understand it as the family of $N+1$ POVMs that have the POVM $\povm{P}$ as its first element, and the $N$ extremal trivial POVMs, in the same order as in Eq.~\eqref{eq:extended-POVM}.

\begin{figure}[t]
    \centering
    \includegraphics[width=10cm]{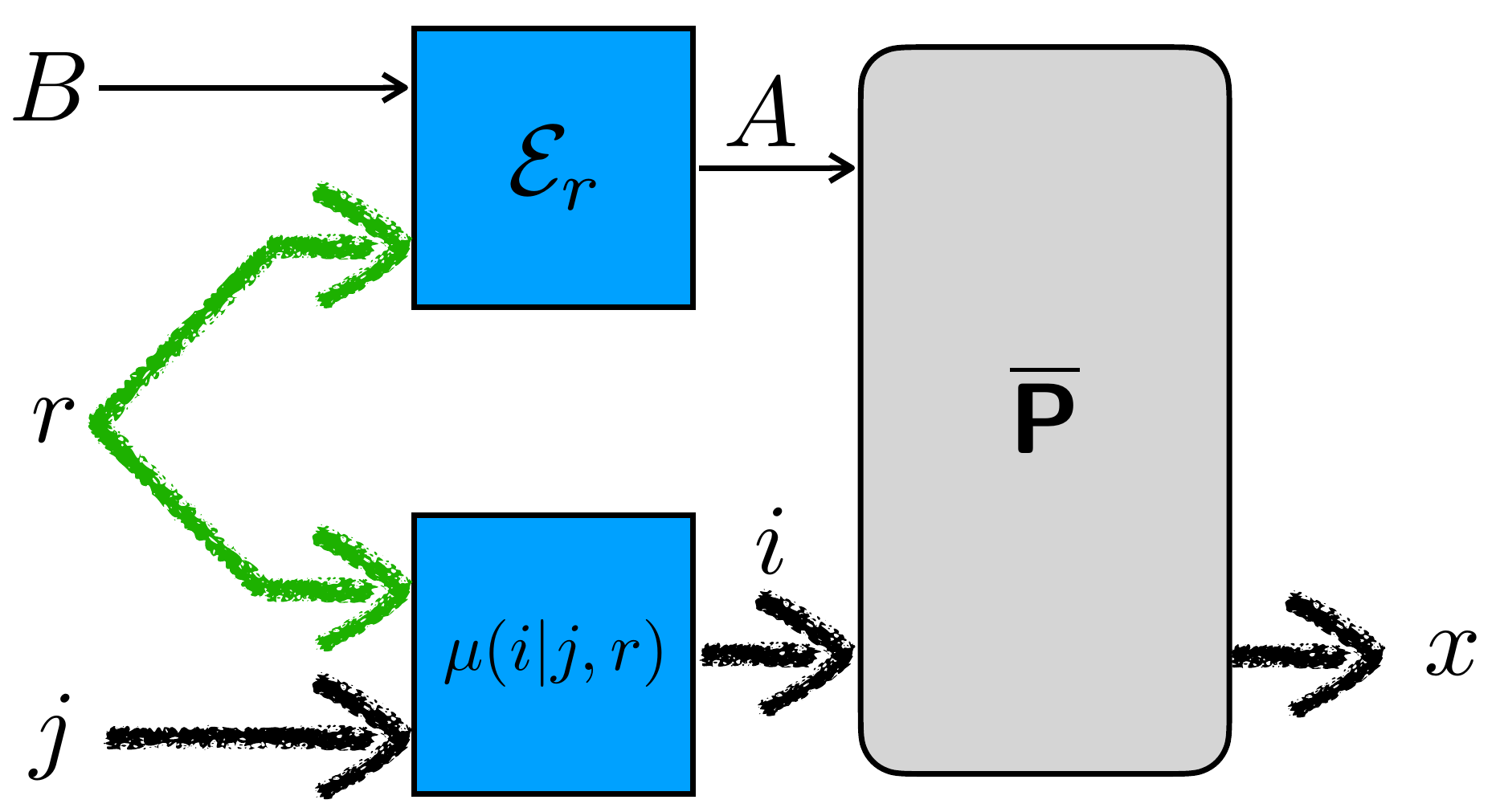}
    \caption{\textbf{Local preprocessing with shared randomness (LPSR).} Given a POVM $\povm{P}$ with outcome set $\set{X}$, we uniquely associate to it a programmable POVM $\overline{\povm{P}}$ (in grey). Correspondingly, any fuzzifying operation on $\povm{P}$ is uniquely associated with an LPSR operation (in blue) on $\overline{\povm{P}}$. The quantum preprocessing channel $B\to A$ is an arbitrary completely positive trace-preserving linear map, while the classical preprocessing on the program alphabet $\set{I}:=\{0\}\cup\set{X}$ is restricted to act as the identity channel on all program values different from zero. Free shared randomness (in green) between the two local preprocessing channels is allowed. In the picture, thin arrows represent quantum systems; thick arrows represent classical systems.}
    \label{fig:figure}
\end{figure}

Since $\overline{\povm{P}}$ is a family of POVMs, following~\cite{buscemi2020complete} we regard it as a \textit{programmable} POVM, where the program is an element $i$ of the set $\set{I}:=\{0\}\cup\set{X}=\{0,1,2,\dots,N\}$. As schematically depicted in Fig.~\ref{fig:figure}, a programmable POVM is a device with two inputs and one output: one quantum input $A$, i.e., the quantum system being measured, one classical input $i\in\set{I}$, i.e. the program deciding which POVM to measure, and one classical output, i.e., the outcome $x\in\set{X}$.

We now want to show that fuzzifying operations on a POVM $\povm{P}$ can be seen as a suitably chosen family of preprocessing channels applied to the corresponding extended object $\overline{\povm{P}}$. Such a family of preprocessing is obtained if the free operations are taken as follows (again, refer to Fig.~\ref{fig:figure} for a diagram):
\begin{enumerate}
    \item any quantum preprocessing channel $\mE$ mapping system $B$ into system $A$ is a free operation;
    \item any classical preprocessing channel (i.e., conditional probability) from $\set{I}$ to $\set{I}$, \textit{acting identically} on $\set{I}\setminus\{0\}$, is a free operation;
    \item shared randomness between the above is free.
\end{enumerate}
For simplicity, let us call preprocessing channels of the above form \textit{local preprocessing with shared randomness}, or LPSR operations. Let us also introduce the notation $\overline{\povm{P}}\succlpsr\overline{\povm{Q}}$ to indicate that the programmable device $\overline{\povm{P}}$ on $A$ can be transformed into the programmable device $\overline{\povm{Q}}$ on $B$ by means of an LPSR operation.

\begin{theorem}\label{th:fuzzifying-preprocessings}
    Given two POVMs $\povm{P}=\{P^x_A\}_{x\in\set{X}}$ and $\povm{Q}=\{Q^x_B\}_{x\in\set{X}}$, possibly defined on different Hilbert spaces $\sH_A$ and $\sH_B$ but with the same outcome set $\set{X}$,
    \begin{align*}
     \povm{P}\succsharp\povm{Q}\quad \iff\quad \overline{\povm{P}}\succlpsr\overline{\povm{Q}} \;.
    \end{align*}
\end{theorem}

\begin{proof}

Let us write explicitly the action of a LPSR operation on $\overline{\povm{P}}:A\times\set{I}\to \set{X}$. By construction, the resulting device $\overline{\povm{Q}}:B\times\set{I}\to \set{X}$ is given by
\begin{enumerate}
    \item $Q_B^{x|0}=\sum_r\nu(r)\mE^\dag_r\left(\sum_{i=0}^N\mu(i|r)P^{x|i}_A\right)$, for all $x\in\set{X}$;
    \item $Q^{x|i}_B=\delta_{x,i}\openone_B$, for all $x\in\set{X}$ and all $i\in\set\{1,\dots,N\}$.
\end{enumerate}
In the above we have used the facts that:
\begin{enumerate}
    \item the probability distribution $\nu(r)$ models the shared randomness between quantum and classical preprocessings;
    \item $\mE^\dag_r:A\to B$ are all completely positive unit-preserving linear maps (i.e., the quantum preprocessings);
    \item the conditional probability $\mu(i|r)$ represents the action of the classical preprocessing channel when the input program value is $0$, i.e., $\mu(i|r)\equiv\mu(i|j=0,r)$;
    \item for the remaining input program values in $\set{I}\setminus\{0\}$ the classical preprocessing acts as the noiseless channel, and thus the trival POVMs corresponding to $i\in\set\{1,\dots,N\}$ are preserved (apart from the change of system $A\to B$).
\end{enumerate}

The proof then simply amounts to show that the equation
\begin{align}
Q_B^{x|0}&=\sum_r\nu(r)\mE^\dag_r\left(\sum_{i=0}^N\mu(i|r)P^{x|i}_A\right)\nonumber\\
&=\sum_r\sum_{i=0}^N\nu(r)\mu(i|r)\mE^\dag_r(P^{x|i}_A)\label{eq:to-change}
\end{align}
can be rewritten in the form of Eq.~\eqref{eq:def-fuzzifying}. This can be done with the following replacements:
\begin{enumerate}
    \item for each $r$, define $\mu_r:=\mu(0|r)$, so that $1-\mu_r:=\sum_{i=1}^N\mu(i|r)$;
    \item for each $r$ such that $1-\mu_r>0$, define $p_r(i):=\frac{1}{1-\mu_r}\mu(i|r)$ for all $i\neq 0$;
    \item for each $r$ such that $1-\mu_r=0$, define $p_r(i)$ as arbitrary positive numbers, for example, $1/N$ for all $i\neq 0$.
\end{enumerate}
With the above substitutions, Eq.~\eqref{eq:to-change} becomes
\begin{align*}
    Q_B^{x|0}&=\sum_r\nu(r)\mE_r^\dag\left(
    \mu_r P^{x|0}_A+(1-\mu_r)\sum_{i\neq 0}p_r(i)T_A^{x|i}
    \right)\\
    &=\sum_r\nu(r)\left\{
    \mu_r\mE_r^\dag(P^{x|0}_A)+(1-\mu_r)\sum_{i\neq 0}p_r(i)\mE_r^\dag(T_A^{x|i})
    \right\}\\
    &=\sum_r\nu(r)\left\{
    \mu_r\mE_r^\dag(P^{x|0}_A)+(1-\mu_r)\sum_{i\neq 0}p_r(i)T_B^{x|i}
    \right\}\\
    &=\sum_r\nu(r)\left\{
    \mu_r\mE_r^\dag(P^{x|0}_A)+(1-\mu_r)p_r(x)\openone_B
    \right\}\;. 
\end{align*}
Since $\sum_r\nu(r)\mu_r\mE^\dag_r$ is a scalar multiple of a completely positive unit-preserving linear map, we can write it is as $\mu\mE^\dag$ with $\mE^\dag$ completely positive and unit-preserving, and $\mu\in[0,1]$. For the same reason, $\sum_r\nu(r)(1-\mu_r)p_r(x)\openone_B$ is proportional to a trivial POVM. Since the POVM on the left-hand side is normalized by definition, we have that the right-hand side can be written as in~\eqref{eq:def-fuzzifying}, as claimed.
\end{proof}

Hence, we see that fuzzifying operations given in Def.~\ref{def:fuzzifying-op}, even though they are not linear in $\povm{P}$, they can nonetheless be regarded as linear maps, more precisely, as LPSR operations, acting on the programmable device $\overline{\povm{P}}$.

\section{Measures of sharpness}\label{sec:ozawa-correlations-and-sharpness}

In what follows, we introduce and study a class of sharpness measures, that are by definition non-increasing under the action of fuzzifying operations and thus provide a class of monotones for a resource theory of sharpness. Our construction generalizes other sharpness measures introduced in~\cite{Liu-Luo-2021-PRA-quantifying-unsharpness,Mitra-2022aa-IJTP-quantifying-unsharpness} and it is based on the theory of EPR-Ozawa measurement correlations.

\subsection{EPR-Ozawa measurement correlations}

In order to clarify in a mathematically rigorous way the meaning of the statement, crucial for the EPR argument~\cite{Einstein1935}, that ``two observables have the same value'', Ozawa  introduced the concept of quantum perfect correlations~\cite{OZAWA-2005-PLA-perfect-correlations,OZAWA-2006-Annals-Quantum-perfect-correlations}.


\begin{definition}
Given a state $\rho_{A}$ on $\sH_A$ and two POVMs on $A$ with the same outcome set $\set{X}$, $\povm{P}=\{P^x_A\}_{x\in\set{X}}$ and $\povm{Z}=\{Z^x_A\}_{x\in\set{X}}$, we say that $\povm{P}$ and $\povm{Z}$ are \emph{jointly distributed} in $\rho$ if and only if $\Tr{P^x_AZ^{x'}_A\ \rho_A}\ge 0$ for all $x,x'\in\set{X}$. If they are jointly distributed, their \emph{degree of correlation} is defined as
\begin{align*}
\kappa_\rho(\povm{P}:\povm{Z}):=\sum_{x\in\set{X}}\Tr{P^x_A Z^x_A\ \rho_{A}}\;.
\end{align*}
\end{definition}

\begin{remark}\label{rem:mitra}
The sharpness measure $\mathcal{P}^L(\rho;\povm{P}):=\sum_x\Tr{\rho_A\ (P^x_A)^2}$ introduced in Eq.~(6) of Ref.~\cite{Mitra-2022aa-IJTP-quantifying-unsharpness}, is a special case of the degree of correlation: more precisely, it coincides with the degree of \emph{autocorrelation} $\kappa_\rho(\povm{P}:\povm{P})$.
\end{remark}

In what follows we will consider in particular the case in which the state $\rho_A$ is maximally mixed. We do this for two reasons. The first is that, in the maximally mixed state, any two POVMs are always jointly distributed, so that their degree of correlation is always defined. In that case, we will use the term \textit{degree of uniform correlations}, together with the short-hand notation
\begin{align}\label{eq:uniform-degree-corr}
\kappa_u(\povm{P}:\povm{Z}):=\frac{1}{d_A}\sum_{x\in\set{X}}\Tr{P^x_A\ Z^x_A}\;,
\end{align}
where the subscript $u$ stands for the ``uniform'', i.e., maximally mixed state $\frac{1}{d_A}\openone_A$. A second reason to focus on the case of the maximally mixed state, is that the degree of uniform correlations can also be written as follows:
\begin{align*}
\kappa_u(\povm{P}:\povm{Z})=\sum_{x\in\set{X}}\Tr{({}^tP^x_{A'}\otimes Z^x_A)\ \ketbra{\Phi^+}_{A'A}}\;,
\end{align*}
where $\ket{\Phi^+}_{A'A}:=\frac{1}{\sqrt{d_A}}\sum_{i=1}^{d_A}\ket{i}_{A'}\otimes\ket{i}_A$ is the maximally entangled state between $A$ and an auxiliary system $A'\cong A$, and the left-hand superscript ${}^t\bullet$ denotes the transposition done with the respect to the basis $\{\ket{i}\}_i$ used in the definition of $\ket{\Phi^+}_{A'A}$. This shows that the degree of uniform correlation $\kappa_u(\povm{P}:\povm{Z})$ can be, in principle, always experimentally estimated, from the joint probability distribution of two POVMs measured by two separated parties sharing the maximally entangled state. 

\subsection{Tuning games, tuning preorder, and sharpness monotones}

When computing the degree of uniform correlation in Eq.~\eqref{eq:uniform-degree-corr}, let us imagine that the POVM $\povm{Z}$ plays the role of a ``reference measurement'', with respect to which the outcome set is fixed. Let us hence consider a reference POVM $\povm{Z}=\{Z^x_R\}_{x\in\set{X}}$, with outcome set $\set{X}$ and defined on some reference system $R$ with Hilbert space $\sH_R$. The choice of a reference POVM can be understood as defining a ``tuning game'', closely related to the notion of ``guessing games'', which have already been widely used in the construction of operational resource monotones: see, e.g.,~\cite{buscemi-datta-strelchuk-2014,buscemi2015degradable,buscemi-datta-2016-divisibility,Skr-Linden-2019,carmeli-prl-2019-incompatibility-witnesses,buscemi2020complete,Carmeli-2022-quantum-guessing-games-review}. In what follows, we utilize tuning games to compare different POVMs with respect to their expected utilities in playing such games.

Fixed a reference POVM $\povm{Z}=\{Z^x_R\}_{x\in\set{X}}$, we can now measure how another POVM, say, $\povm{P}=\{P^x_A\}_{x\in\set{X}}$, with the same outcome set of the reference but otherwise arbitrary, can be ``tuned'' with the reference $\povm{Z}$. Here, we focus on the following quantity
\begin{align}\label{eq:synchro-payoff}
\kappa_u^*(\povm{P}\|\povm{Z}):&=\max_{\mL}\kappa_u(\mL(\povm{P}):\povm{Z})\\
&=\max_{\mL}\frac{1}{d_R}\sum_{x\in\set{X}}\Tr{\mL(P^x_A)\ Z^x_R}\nonumber\;,
\end{align}
where the optimization is done over all fuzzifying operations $\mL$, as given in Eq.~\eqref{eq:def-fuzzifying}. In other words, the quantity $\kappa_u^*(\povm{P}\|\povm{Z})$ measures the degree of uniform correlations that can be established between a POVM $\povm{P}$ (seen here as the resource that the player employs to play the tuning game) and the reference $\povm{Z}$, by means of a sharpness-non-increasing operation applied on $\povm{P}$. We will refer to the quantity $\kappa^*_u(\povm{P}\|\povm{Z})$ as the \emph{tuning degree} of $\povm{P}$ with respect to $\povm{Z}$. Notice that while the degree of correlation~\eqref{eq:uniform-degree-corr} is symmetric in the POVMs, the tuning degree~\eqref{eq:synchro-payoff} is not, since the optimization is done only on one of the two POVMs. The notation $\kappa_u^*(\povm{P}\|\povm{Z})$ reflects this. Notice also that other choices for the tuning process may be done: this freedom is similar to what happens, for example, in the resource theory of entanglement, for which there exist different, though all operationally meaningful, notions of entanglement manipulation, such as LOCC~\cite{BBP+96} or LOSR~\cite{buscemi2012all}. However, in the context of the present paper it is natural to define the optimization with respect to the same class of transformations that is used to define the sharpness preorder $\succsharp$ in Definition~\ref{def:sharpness-preorder}.

From the definition~\eqref{eq:synchro-payoff}, it is clear that a sharp POVM $\povm{P}$ allows for \textit{ideal} tuning. Since, as Theorem~\ref{th:pre-clean} states, sharp POVMs are exactly those that can be transformed into \textit{any other} POVM by means of a suitable fuzzifying operation, the quantity $\kappa_u^*(\povm{P}\|\povm{Z})$, if $\povm{P}$ is sharp, can be pushed up to its maximum value, namely,
\begin{align}\label{eq:guessing-prob}
\kappa_u^*(\povm{Z}):=\max_{\{\tilde{Z}^x_R\}_x:\text{ POVM}} \kappa_u(\tilde{\povm{Z}}:\povm{Z})=\max_{\{\tilde{Z}^x_R\}_x:\text{ POVM}}\frac{1}{d_R}\sum_{x\in\set{X}}\Tr{\tilde{Z}^x_R\ Z^x_R}\;,
\end{align}
which is a quantity that only depends on the reference $\povm{Z}$.

More generally, for any given reference $\povm{Z}$, the tuning degree $\kappa_u^*(\povm{P}\|\povm{Z})$, seen as a function of $\povm{P}$, constitutes a \textit{sharpness monotone}, i.e., a function that is maximal on sharp POVMs and is by definition non-increasing under the action of sharpness-non-increasing operations.

\begin{remark} 
Notice that the quantity appearing in Eq.~\eqref{eq:guessing-prob} is equal to the maximum probability of correctly discriminating among the states of the ensemble $\{p(x),\rho(x)\}_{x\in\set{X}}$, where $p(x):=\frac{1}{d_R}\Tr{Z^x_R}$ and $\rho^x_R:=\frac{1}{\Tr{Z^x_R}}Z^x_R$. It can thus be effectively computed using a simple semi-definite program, see for example Section~3.1.2 of Ref.~\cite{watrous2018theory}.
\end{remark}

\begin{remark}
Let us go back for a moment to the degree of autocorrelation introduced by Mitra in~\cite{Mitra-2022aa-IJTP-quantifying-unsharpness}, and already considered in Remark~\ref{rem:mitra}. It is easy to verify that it is not a sharpness monotone. This is a consequence of the fact that $\kappa_\rho(\povm{P}:\povm{P})$ can always be made equal to its algebraic maximum (i.e., equal to one), simply by replacing the initial POVM $\povm{P}$ with any one of the trivial POVMs $\povm{T}^{(x)}$, $x\in\set{X}$, introduced in~\eqref{eq:extended-POVM}, an action that is by definition a fuzzifying operation. One way around this problem is to use the \textit{degree of autotuning} $\kappa_u^*(\povm{P}\|\povm{P})$, which \textit{is} a sharpness monotone instead.
A relation between the degree of autotuning $\kappa_u^*(\povm{P}\|\povm{P})$ (a sharpness monotone) and the degree of uniform autocorrelation $\kappa_u(\povm{P}:\povm{P})$ (not a sharpness monotone) can be found using the theory of \emph{pretty good measurements}~\cite{belavkin-1975-pretty-good,barnum-knill}, using which we can show that
\begin{align*}
\kappa_u^*(\povm{P}\|\povm{P})\ge\kappa_u(\povm{P}:\povm{P})\ge\kappa_u^*(\povm{P}\|\povm{P})-(1-\kappa_u^*(\povm{P}\|\povm{P}))\;.
\end{align*}
In other words, if the degree of autotuning $\kappa_u^*(\povm{P}\|\povm{P})$ is not smaller than $1-\epsilon$, the degree of uniform autocorrelation $\kappa_u(\povm{P}:\povm{P})$ is not smaller than $1-2\epsilon$: i.e., it provides a ``pretty good'' estimate of $\kappa_u^*(\povm{P}\|\povm{P})$, even though it is not optimized over all fuzzifying operations.
\end{remark}


We now use the operational task of tuning to compare two POVMs with the same outcome set as follows.

\begin{definition}[tuning preorder]
Given a tuning game as a reference POVM $\povm{Z}=\{Z^x_R\}_{x\in\set{X}}$ and two POVMs $\povm{P}=\{P^x_A\}_{x\in\set{X}}$ and $\povm{Q}=\{Q^x_B\}_{x\in\set{X}}$, possibly defined on different Hilbert spaces $\sH_A$ and $\sH_B$ but with the same outcome set as the reference $\povm{Z}$, we say that $\povm{P}$ is \emph{more tunable} than $\povm{Q}$ with respect to $\povm{Z}$, and write
\begin{align}\label{eq:stronger-cond}
\povm{P}\succeq_{\povm{Z}}^{\operatorname{t}}\povm{Q}\;,
\end{align}
whenever $\kappa_u^*(\povm{P}\|\povm{Z})\ge \kappa_u^*(\povm{Q}\|\povm{Z})$.

Further, given two POVMs $\povm{P}=\{P^x_A\}_{x\in\set{X}}$ and $\povm{Q}=\{Q^x_B\}_{x\in\set{X}}$, possibly defined on different Hilbert spaces $\sH_A$ and $\sH_B$ but with the same outcome set $\set{X}$, we say that $\povm{P}$ is \emph{always more tunable} than $\povm{Q}$, and write
\begin{align}
\povm{P}\succcorr\povm{Q}\;,
\end{align}
whenever $\povm{P}\succeq_{\povm{Z}}^{\operatorname{t}}\povm{Q}$ for all reference POVMs $\povm{Z}$ with outcome set $\set{X}$.
\end{definition}

\subsection{Robustness of tunability}\label{subset:robustness}

By looking at the ``tuning advantage'' that a given POVM $\povm{P}$ has with respect to trivial POVMs, we introduce the following definition:

\begin{definition}[tunability robustness]\label{def:robust}
For any POVM $\povm{P}_{A} = \{P^{x}_{A}\}_{x \in \set{X}}$ with outcome set $\set{X}$, its \emph{tunability robustness} $\mRtun(\povm{P})$ is defined by
\begin{equation}\label{eq:tunability-robustness}
 1 + \mRtun(\povm{P}):=\max_{\povm{Z}}\frac{\kappa_u^*(\povm{P}\|\povm{Z})}{\max_{\povm{T}}\kappa_u^*(\povm{T}\|\povm{Z})}, 
\end{equation}
where the first maximization is over all reference POVM $\povm{Z}=\{Z^x_R\}_{x\in\set{X}}$, and the second maximization is done over all trivial POVMs  $\povm{T}=\{T^x_A\}_{x\in\set{X}}$. 
\end{definition}

The robustness of tunability satisfies the three basic properties of a robustness measure, that is, convexity, monotonicity, and faithfulness. We postpone the proof of this fact after Theorem~\ref{th:main} in the next section. What we show here is that the tunability robustness is never larger than the measurement robustness~\cite{Uola-etal-2019, Skr-Linden-2019, Oszmaniec2019operational, takagi2019general}, which is defined as
\begin{equation}
\mR(\povm{P}) := \min \left\{ r \geq 0\ \left|\ \left\{\frac{P^{x}_{A} + rQ^{x}_{A}}{1 + r}\right\}_{x \in \set{X}} \in \mathbb{T}^{\set{X}}\ :\  \povm{Q} = \left\{ Q^{x}_{A} \right\} \text{ is POVM} \right\}\right., \label{eq:robustness-of-povm}
\end{equation}
where $\mathbb{T}^{\set{X}}$ denotes the set of trivial POVMs on $A$ with outcome set $\set{X}$.

This fact can be shows as follows. From the definition of the measurement robustness~\eqref{eq:robustness-of-povm}, for any POVM $\povm{P}= \{P^{x}_{A}\}_{x \in \set{X}}$, there exists a POVM $\povm{Q}= \{Q^{x}_{A}\}_{x \in \set{X}}$ and a probability distribution $\{q(x)\}_{x \in \set{X}}$ such that $\forall x$, $P^{x}_{A} = [1 + \mR(\povm{P})]q(x)\openone_{A} - \mR(\povm{P})Q^{x}_{A}$. Then, for any reference system $R$ and any POVM $\povm{Z} = \{Z^x_R\}_{x \in \set{X}}$, we have
\begin{align*}
d_R\ \kappa_u^*(\povm{P}\|\povm{Z})&= \max_{\mL} \sum_{x \in \set{X}}\Tr{\mL(P^{x}_{A})\  Z^x_R} \\
&\equiv \sum_{x \in \set{X}}\Tr{\mL^*(P^{x}_{A})\  Z^x_R} \\
&= \sum_{x \in \set{X}}\Tr{\mL^*\left\{[1 + \mR(\povm{P})]q(x)\openone_{A} - \mR(\povm{P})Q^{x}_{A}\right\}\  Z^x_R} \\
&\le \sum_{x \in \set{X}}\Tr{\mL^*\left\{[1 + \mR(\povm{P})]q(x)\openone_{A}\right\}\  Z^x_R} \\
&= [1 + \mR(\povm{P})]\sum_{x \in \set{X}}\Tr{\mL^*\left\{q(x)\openone_{A}\right\}\  Z^x_R}\\
&\le [1 + \mR(\povm{P})]\max_{\povm{T}\in\mathbb{T}}\sum_{x \in \set{X}}\Tr{T^x_R\  Z^x_R}\\
&= d_R\ [1 + \mR(\povm{P})]\max_{\povm{T}\in\mathbb{T}}\kappa_u^*(\povm{T}\|\povm{Z})\;, 
\end{align*}
where $\mL^{\ast}$ in the second equality is the optimal fuzzifying operation; the first inequality is obtained simply discarding the negative term; the second inequality come from the fact that $\mL^*$ acting on a trivial POVM is again a trivial POVM, so that, maximizing over all trivial POVMs can only achieve a better score.

Summarizing, the above shows that, for any reference system $R$ and any POVM $\povm{Z} = \{Z^x_R\}_{x \in \set{X}}$,
\begin{equation*}
    \frac{\kappa_u^*(\povm{P}\|\povm{Z})}{\max_{\povm{T}}\kappa_u^*(\povm{T}\|\povm{Z})}\le 1+\mR(\povm{P})\;.
\end{equation*}
Optimizing the left-hand side over $\povm{Z}$ (the right-hand side does not depend on $\povm{Z}$), we finally obtain
\begin{equation}\label{eq:robustness-bound}
    \mRtun(\povm{P})\le \mR(\povm{P})\;,
\end{equation}
as claimed.

\section{Equivalence of comparisons}\label{sec:blackwell}

In this section we develop the theory of statistical comparison for the sharpness preorder that we introduced above. Statistical comparison is a concept introduced by Blackwell~\cite{blackwell1953} with the aim of extending the ideas of Lorenz curves and majorization~\cite{hardy1952inequalities,MarshallOlkin} to more general scenarios. It establishes an equivalence between two kinds of preorders: a ``sufficiency'' preorder, analogous to the majorization preorder, which is given by the existence of a suitable transformation (e.g., a doubly stochastic matrix, in the case of majorization) between two objects; and a ``game-theoretic'' preorder, which instead concerns the comparison of the expected performance with respect to a certain class of statistical tests (e.g., hypothesis testing, in the case of Lorenz curves). Such an equivalence between, on the one hand, the existence of a transformation and, on the other hand, the comparison of operational utilities (i.e., the ``monotones'' of resource theories), summarizes the core concept that lies at the basis of all resource theories~\cite{Chitambar-Gour2019resource-theories}. In this spirit, various generalizations of Blackwell's theory of statistical comparison~\cite{buscemi-CMP-2012,Buscemi-ProbInfTrans-2016,jencova-comparison-2015,buscemi-gour-2017-q-lorenz,buscemi2018reverse-data-proc,Buscemi-sutter-tomamichel-2019information,jencova-comparison-2021} have been successfully applied in several specific resource-theoretic scenarios, including entanglement and nonlocality theory~\cite{buscemi2012all,Schmid2020typeindependent,Zhou_2020,rosset-schmid-buscemi-2020-prl}, quantum communication theory~\cite{buscemi-datta-strelchuk-2014,Rosset-Buscemi-Liang-PRX}, open quantum systems dynamics~\cite{buscemi-NCPTP,buscemi-datta-2016-divisibility}, quantum coherence~\cite{regula-buscemi-2020-coherence-prr}, quantum thermodynamics~\cite{buscemi-2015-fully-quantum-second-laws,Gour:2018aa-jennings-buscemi-2018}, and quantum measurement theory~\cite{Buscemi-ProbInfTrans-2016,Skr-Linden-2019,buscemi2020complete,ji2021incompatibility,buscemi-2022-unifying-instrument-incompatibility}.

The Blackwell--like theorem that we prove in this work is the following.

\begin{theorem}\label{th:main} 
Given two POVMs $\povm{P}=\{P^x_A\}_{x\in\set{X}}$ and $\povm{Q}=\{Q^x_B\}_{x\in\set{X}}$, possibly defined on different Hilbert spaces $\sH_A$ and $\sH_B$ but with the same outcome set $\set{X}$, $\povm{P}$ can be transformed into $\povm{Q}$ by means of a fuzzifying operation, that is,
\begin{align*}
\povm{P}\succsharp\povm{Q}\qquad(\iff\overline{\povm{P}}\succlpsr\overline{\povm{Q}})
\end{align*}
if and only if $\povm{P}$ is always more tunable than $\povm{Q}$, that is,
\begin{align}\label{eq:equivalent}
\povm{P}\succcorr\povm{Q}\;.
\end{align}
Moreover, the preorder~\eqref{eq:equivalent} can be restricted without loss of generality to reference POVMs defined on the same Hilbert space as $\povm{Q}$, i.e., $\sH_B$.

Hence, the tuning degrees $\kappa^*_u(\povm{P}\|\povm{Z})$, for varying reference POVM $\povm{Z}$, provide a complete set of monotones for the resource theory of sharpness.
\end{theorem}

\begin{proof}[Proof of Theorem~\ref{th:main}]
Our aim is to show that the condition about the existence of a fuzzifying operation transforming $\povm{P}$ into $\povm{Q}$, can be equivalently written as Eq.~\eqref{eq:equivalent}.

From Theorem~\ref{th:fuzzifying-preprocessings}, we know that $\povm{P}\succsharp\povm{Q}$, if and only if there exists a LPSR operation transforming the extended programmable device corresponding to $\povm{P}$, i.e., $\overline{\povm{P}}$, into the extended programmable device corresponding to $\povm{Q}$, i.e., $\overline{\povm{Q}}$. For notational convenience, let us denote the LPSR operation as $\mL$ and the elements of the resulting programmable device as $\mL(\overline{\povm{P}})^{x|i}_B$.

Let us now consider an arbitrary but fixed complete set of density matrices $\{\gamma_B^b\}_{b\in\set{B}}$, in the sense that the linear span of $\{\gamma_B^b\}_{b\in\set{B}}$ coincides with the set of all linear operators on $\sH_B$. Then, $\povm{P}\succsharp\povm{Q}$ if and only if
\begin{align}\label{eq:equivalent2}
\Tr{\mL(\overline{\povm{P}})^{x|i}_B\ \gamma_B^b}=\Tr{Q^{x|i}_B\ \gamma_B^b}\;,\qquad\forall x,\forall i,\forall b\;.
\end{align}
Looking at the two conditional distributions above as vectors in $\mathbb{R}^{|\set{X}|\times|\set{I}|\times|\set{B}|}$, that is, $\vecp_\mL$ and $\vecq$, respectively, let us consider the subset of $\mathbb{R}^{|\set{X}|\times|\set{I}|\times|\set{B}|}$ defined as
\begin{align*}
\mathcal{C}(\overline{\povm{P}}):=\left\{\vecp_\mL: p_\mL(x|i,b)=\Tr{\mL(\overline{\povm{P}})^{x|i}_B\ \gamma_B^b}\right\}\;,
\end{align*}
where $\mL$ can range over all LPSR operations.
Then, Eq.~\eqref{eq:equivalent2} can be equivalently rewritten as
\begin{align}\label{eq:equivalent5}
\vecq\in\mathcal{C}(\overline{\povm{P}})\;.
\end{align}

The crucial observation now is that, since the definition of LPSR operations involves free shared randomness, they form a convex set. For this reason, also $\mathcal{C}(\overline{\povm{P}})$ is a convex subset of $\mathbb{R}^{|\set{X}|\times|\set{I}|\times|\set{B}|}$. Hence, as a consequence of the separation theorem for convex sets, we can rewrite condition~\eqref{eq:equivalent5} in terms of linear functionals as follows
\begin{align*}
\boldsymbol{\lambda}\cdot\vecq\le\max_{\vecp_\mL\in\mathcal{C}(\overline{\povm{P}})}\boldsymbol{\lambda}\cdot\vecp_\mL\;,\qquad\forall\boldsymbol{\lambda}\in\mathbb{R}^{|\set{X}|\times|\set{I}|\times|\set{B}|}\;,
\end{align*}
which, once rewritten in a more explicit form, becomes
\begin{align*}
\max_{\mL}\sum_{x,i,b}\lambda_{xib}\Tr{\mL(\overline{\povm{P}})^{x|i}_B\ \gamma_B^b}\ge\sum_{x,i,b}\lambda_{xib}\Tr{Q^{x|i}_B\ \gamma_B^b}\;,\qquad\forall \lambda_{xib}\in\mathbb{R}\;.
\end{align*}
Introducing the self-adjoint operators $\Gamma^{xi}_B:=\sum_b\lambda_{xib}\gamma^b_B$, the above condition becomes
\begin{align*}
\max_{\mL}\sum_{x,i}\Tr{\mL(\overline{\povm{P}})^{x|i}_B\ \Gamma_B^{xi}}\ge\sum_{x,i}\Tr{Q^{x|i}_B\ \Gamma^{xi}_B}\;,\qquad\forall\text{ self-adjoint }\{\Gamma^{xi}_B\}_{x,i}\;.
\end{align*}

First, we notice that since, by construction, $\mL(\overline{\povm{P}})^{x|i}_B=Q^{x|i}_B=\delta_{x,i}\openone_B$ for all $x,i\in\set{X}$ and any choice of the LPSR operation $\mL$, we can in fact focus only on the case $i=0$. Therefore, in what follows, we will only consider the conditions
\begin{align}\label{eq:equivalent3}
\max_{\mL}\sum_{x}\Tr{\mL(\overline{\povm{P}})^{x|0}_B\ \Gamma_B^{x}}\ge\sum_{x}\Tr{Q^x_B\ \Gamma^{x}_B}\;,\qquad\forall\text{ self-adjoint }\{\Gamma^{x}_B\}_{x}\;.
\end{align}

The next step is to notice that it is possible to shift and rescale the operators $\Gamma^{x}_B$ in such a way that, without loss of generality, we can restrict condition~\eqref{eq:equivalent3} to families of operators $\{Z^x_B\}$ such that $\sum_xZ^x_B=\openone_B$ and $Z^x_B\ge 0$ for all $x\in\set{X}$, i.e., POVMs $\povm{Z}=\{Z^x_B\}_{x\in\set{X}}$ on $B$. In fact, by the linearity of the trace and the fact that $\sum_x\mL(\overline{\povm{P}})^{x|0}_B=\sum_xQ^x_B=\openone_B$, for any choice of self-adjoint operators $\{\Gamma^{x}_{B}\}_{x}$ in \eqref{eq:equivalent3}, we have 
\begin{align*}
    \max_{\mL}\sum_{x}\Tr{\mL(\overline{\povm{P}})^{x|0}_B\ \Gamma_B^{x}}&\ge\sum_{x}\Tr{Q^x_B\ \Gamma^{x}_B}\\
    &\Big\Updownarrow\\
  \max_{\mL}\sum_{x}\Tr{\mL(\overline{\povm{P}})^{x|0}_{B}\ \frac{\Gamma^{x}_{B} + \alpha \Lambda_{B}}{\beta}} &\ge \sum_{x}\Tr{Q^{x}_{B}\ \frac{\Gamma^{x}_{B} + \alpha \Lambda_{B}}{\beta}},
\end{align*}
for any choice of $\alpha \in \op{R}$, $\beta > 0$, and $\Lambda_{B}$ linear operator on $\sH_{B}$. In particular, by taking $\alpha=1$ and $\Lambda_B=\frac{1}{|\set{X}|}(\beta\openone-\sum_x\Gamma^x_B)$ we obtain
\begin{align*}
    \frac{\Gamma^{x}_{B} + \alpha \Lambda_{B}}{\beta}=\frac{\openone}{|\set{X}|}+\frac{1}{\beta}\left(\Gamma^x_B-\frac{1}{|\set{X}|}\sum_y\Gamma^y_B\right)\;,
\end{align*}
so that, by choosing $\beta>0$ sufficiently large, it is always possible to make the above operators the elements of a POVM\footnote{In fact, by taking $\beta>0$ large enough, the obtained POVM has elements which are all invertible. On this point, see Remark~\ref{rem:special-cases}}.

In this way, we can rewrite condition~\eqref{eq:equivalent3}, which we recall is equivalent to $\povm{P}\succsharp\povm{Q}$, as follows:
\begin{align*}
\max_{\mL}\sum_{x}\Tr{\mL(\overline{\povm{P}})^{x|0}_B\ Z^x_B}\ge \sum_{x}\Tr{Q^x_B\ Z^x_B}\;,\qquad\forall\text{ POVMs }\povm{Z}=\{Z^x_B\}_x\;,
\end{align*}
namely, 
\begin{align}\label{eq:equvalent4}
\kappa_u^*(\povm{P}\|\povm{Z})\ge \kappa_u(\povm{Q}:\povm{Z})\;,\qquad\forall\text{ POVMs }\povm{Z}=\{Z^x_B\}_x\;.
\end{align}
Finally, since the inequality $\kappa_u^*(\povm{Q}\|\povm{Z})\ge \kappa_u(\povm{Q}:\povm{Z})$ is true by definition, we reach the conclusion that if condition~\eqref{eq:equivalent} holds, then also condition~\eqref{eq:equvalent4} holds, which in turn is equivalent to~\eqref{eq:equivalent2}. Hence, we have proved that if $\povm{P}\succcorr\povm{Q}$ then $\povm{P}\succsharp\povm{Q}$.

The converse is trivial: if $\povm{P}\succsharp\povm{Q}$ then obviously any tuning degree that can be achieved with $\povm{Q}$ can also be achieved with $\povm{P}$, simply because the latter can be transformed into the former, and the compositions of fuzzifying operations is again a fuzzifying operation.
\end{proof}

\begin{remark}\label{rem:special-cases}
The proof of Theorem~\ref{th:main} shows that another, at first sight weaker condition is in fact equivalent to~\eqref{eq:equivalent}, i.e.,
\begin{align}\label{eq:weaker-cond}
\kappa_u^*(\povm{P}\|\povm{Z})\ge \kappa_u(\povm{Q}:\povm{Z})\;,
\end{align}
for all reference POVMs $\povm{Z}=\{Z^x_B\}_{x\in\set{X}}$. Notice that the right-hand side in the above equation is \emph{not} optimized, so \emph{for any given} reference POVM $\povm{Z}$, condition~\eqref{eq:weaker-cond} is in general strictly weaker (i.e., easier to be satisfied) than~\eqref{eq:stronger-cond}. However, it turns out that condition~\eqref{eq:weaker-cond} holds \emph{for all} reference POVMs $\povm{Z}$ if and only if~\eqref{eq:stronger-cond} does. The proof also shows that the reference POVMs $\povm{Z}$ defining the monotones can be restricted, without loss of generality, to POVMs with full-rank elements, i.e., $Z^x_R>0$ for all $x$. This observation may be helpful when performing numerical experiments.
\end{remark}

Since all sharp POVMs and all trivial POVMs are equivalent under sharpness-non-increasing operations, we immediately obtain the following:

\begin{corollary}\label{eq:corollary}
All sharp POVMs achieve exactly the same tuning degree for any reference POVM. The same holds for all trivial POVMs. Hence, for any reference POVM $\povm{Z}=\{Z^x_R\}_{x\in\set{X}}$ and any POVM $\povm{P}=\{P^x_A\}_{x\in\set{X}}$,
\begin{align*}
\frac{1}{d_R}\max_{x}\Tr{Z^x_R}\le\kappa^*_u(\povm{P}\|\povm{Z})\le \kappa^*_u(\povm{Z})\;.
\end{align*}
In particular, any value $\kappa^*_u(\povm{P}\|\povm{Z})$ strictly larger than the trivial lower bound provides a \emph{measurement device-independent witness} of the non-triviality of $\povm{P}$.
\end{corollary}

\subsection{Properties of the tunability robustness}

With Theorem~\ref{th:main} and Corollary~\ref{eq:corollary} at hand, it is easy to prove the properties of the tunability robustness that we anticipated in Subsection~\ref{subset:robustness}.

\begin{theorem}\label{th:tunability-robustness}
    The tunability robustness $\mRtun$ satisfies the following properties:
    \begin{enumerate}
        \item convexity, i.e., given two POVMs with the same outcome set $\set{X}$, $\mRtun(p\povm{P}_1+(1-p)\povm{P}_2)\le p\mRtun(\povm{P})+(1-p) \mRtun(\povm{P}_2)$, for all $p\in[0,1]$;
        \item monotonicity, i.e., for any fuzzifying operation $\mL$, $\mRtun(\mL(\povm{P}))\le \mRtun(\povm{P})$;
        \item faithfulness, i.e., $\povm{P}$ is trivial if and only if $\mRtun(\povm{P})=0$, and $\povm{P}$ is sharp if and only if $\mRtun(\povm{P})=|\set{X}|-1$.
    \end{enumerate}
\end{theorem}

\begin{proof}
    Convexity and monotonicity can be easily proved from the definition~\eqref{eq:tunability-robustness}. We show only faithfulness.

    First of all, if $\povm{P}$ is trivial, then its measurement robustness is zero, i.e., $\mR(\povm{P})=0$~\cite{Uola-etal-2019, Skr-Linden-2019, Oszmaniec2019operational, takagi2019general}, and by Eq.~\eqref{eq:robustness-bound}, also $\mRtun(\povm{P})=0$. Vice versa, if $\mRtun(\povm{P})=0$, then we know that $\povm{P}$ does not provide any advantage in any tuning game with respect to trivial POVMs. But since Theorem~\ref{th:main} shows that there always exists a tuning game that separate non-trivial POVMs from trivial ones, if $\mRtun(\povm{P})=0$, then the POVM $\povm{P}$ must be trivial.

    To prove the case of sharp POVMs, let us first notice that, as a consequence of Corollary~\ref{eq:corollary}, 
    \begin{align*}
        1+\mRtun(\povm{P})&=\max_{\povm{Z}}\max_{\mL}\frac{\sum_x\Tr{\mL(P^x_A)\ Z^x_R}}{\max_x\Tr{Z^x_R}}\\
        &\le\max_{\substack{\zeta^x_R\ge 0 \\ \Tr{\zeta^x_R}\le 1}}\max_{\mL}\sum_x\Tr{\mL(P^x_A)\ \zeta^x_R}\\
        &=\max_{\mL}\sum_x\lambda_{\max}(\mL(P^x_A))\\
        &\le |\set{X}|\;,
    \end{align*}
    where $\lambda_{\max}(X)$ denotes the maximum eigenvalue of self-adjoint operator $X$. The last inequality is a consequence of the fact that the output of a fuzzifying operation is still a POVM so that the maximum eigenvalues of its elements are all upper bounded by 1. Now, a sharp POVM saturates the bound, with $\mL=\id$. Hence, if $\povm{P}$ is sharp, then $\mRtun(\povm{P})=|\set{X}|-1$.

    Conversely, assume that $\povm{P}$ is not sharp. Then, for any fuzzifying operation $\mL$, also $\mL(\povm{P})$ is not sharp, and therefore $\sum_x\lambda_{\max}(\mL(P^x_A))<|\set{X}|$, that is, $\mRtun(\povm{P})<|\set{X}|-1$.
\end{proof}

\begin{remark}
    The optimization over $\povm{Z}$ in the definition of the tunability robustness~\eqref{eq:tunability-robustness} includes, in principle, also the optimization over all finite-dimensional reference systems $R$. From the above proof, however, it turns out that it is enough to restrict to $R\cong A$.
\end{remark}

\section{Summary of the theory}\label{sec:conclusion}

For the reader's convenience, we summarize the main points of the resource theory of sharpness that we have derived.
\begin{itemize}
\item The \emph{objects} of the theory are POVMs. In particular, this means that our resource theory of sharpness does not depend on the specific numerical values associated with each measurement outcome, i.e., the observable's eigenvalues, nor on any particular instrument or measurement process used to realize the POVM.
\item The \emph{free operations} are given by the class of fuzzifying operations, which is by construction convex and closed under sequential composition (see Definition~\ref{def:fuzzifying-op}). Though fuzzifying operations are neither quantum preprocessings nor classical postprocessings of the POVM alone, they can be seen as suitable preprocessings (LPSR operations) of a programmable measurement device that extends the given POVM in a one-to-one way.
\item The \emph{maximal objects} in the resource theory of sharpness (i.e., sharp POVMs) are all equivalent, in the sense that any sharp POVM can be freely transformed into any other sharp POVM with the same outcome set. We recall that, in this work, we define sharp POVMs as those whose elements all possess at least one eigenvector associated with eigenvalue 1 (see Definition~\ref{def:sharp-POVMs}).
\item The \emph{minimal objects} are trivial POVMs, i.e., POVMs whose elements are all proportional (including the possibility of zero elements) to the identity operator. As it happens for sharp POVMs, also trivial POVMs are, as one would expect, all equivalent.
\item The \emph{sharpness monotones} are given by the tuning degrees $\kappa^*_u(\povm{P}\|\povm{Z})$, for varying reference POVM $\povm{Z}$, defined in Eq.~\eqref{eq:synchro-payoff}. A robustness-like measure, i.e., the \textit{tunability robustness}, can be introduced, which is convex, monotone, and faithful, and it is upper bounded by the measurement robustness.
\item A \emph{Blackwell--like theorem} for sharpness holds, i.e., a POVM can be transformed into another POVM by a fuzzifying operation, if and only if there exists no tuning degree for the latter that is higher than for the former. This automatically implies that all sharp POVMs and all trivial POVMs achieve exactly the same tuning degree for any reference POVM, as given in Corollary~\ref{eq:corollary}. By normalizing these two numbers, our sharpness monotones satisfy Busch's requirements for sound sharpness measures~\cite{Busch-2009a-on-the-sharpness-and-bias}. Moreover, sharpness monotones can be used to witness in a \emph{measurement device-independent way} the non-triviality of a POVM, exactly in the same way that semiquantum games can witness non-separability~\cite{buscemi2012all,Branciard2013}.
\end{itemize}


\section*{Acknowledgements}
F.B. acknowledges support from MEXT Quantum Leap Flagship Program (MEXT QLEAP) Grant No.~JPMXS0120319794, from MEXT-JSPS Grant-in-Aid for Transformative Research Areas (A) ``Extreme Universe'' No.~21H05183, and from JSPS KAKENHI, Grants No.~20K03746 and No.~23K03230.
S.M. would like to take this opportunity to thank the ``Nagoya University Interdisciplinary Frontier Fellowship'' supported by Nagoya University and JST, the establishment of university fellowships towards the creation of science technology innovation, Grant Number JPMJFS2120.

\bibliographystyle{unsrturl}
\bibliography{library}

\end{document}